\documentclass[letter]{IEEEtran}
\IEEEoverridecommandlockouts
\usepackage{cite}
\usepackage{amsmath,amssymb,amsfonts}
\usepackage{amssymb}
\usepackage{algpseudocode}
\usepackage{amsfonts}
\usepackage{graphicx}
\usepackage{fancyhdr}  %
\usepackage{cases}
\usepackage{textcomp}
\usepackage{extarrows}
\usepackage{algorithm,algpseudocode}
\usepackage{multirow}
\usepackage{booktabs}
\usepackage{caption}
\usepackage{subcaption}

\usepackage{multirow,tabularx}
\usepackage{mathtools}
\usepackage{xcolor}
\usepackage[english]{babel}
\usepackage{amsthm}
\usepackage[left]{lineno}      %[left]代表编号在左侧

\usepackage{caption}
\def\BibTeX{{\rm B\kern-.05em{\sc i\kern-.025em b}\kern-.08em
    T\kern-.1667em\lower.7ex\hbox{E}\kern-.125emX}}
\begin{document}
\bstctlcite{BSTcontrol}
\title{Beamforming Inferring by Conditional WGAN-GP for Holographic Antenna Arrays}

\author{Fenghao Zhu, Xinquan Wang, Chongwen Huang, Ahmed Alhammadi, Hui Chen \\ Zhaoyang Zhang, Chau~Yuen,~\IEEEmembership{Fellow,~IEEE}, and M\'{e}rouane Debbah,~\IEEEmembership{Fellow,~IEEE}

\thanks{F. Zhu, X. Wang, C. Huang, and Z. Zhang are with College of Information Science and Electronic Engineering, Zhejiang University, Hangzhou 310027, China, and F. Zhu, and C. Huang are also with the State Key Laboratory of Integrated Service Networks, Xidian University, Xi’an 710071, China, and Zhejiang Provincial Key Laboratory of Info. Proc., Commun. \& Netw. (IPCAN), Hangzhou 310027, China (E-mail: chongwenhuang@zju.edu.cn). (Corresponding author: Chongwen Huang).}

\thanks{A. Alhammadi and H. Chen are with Technology Innovation Institute, 9639 Masdar City, Abu Dhabi, UAE (E-mails: Ahmed.Alhammadi@tii.ae; hui.chen@tii.ae).}

\thanks{C. Yuen is with the School of Electrical and Electronics Engineering, Nanyang Technological University, Singapore 639798 (E-mail: chau.yuen@ntu.edu.sg).}

\thanks{M. Debbah is with KU 6G Research Center, Khalifa University of Science and Technology, P O Box 127788, Abu Dhabi, UAE and CentraleSupelec, University Paris-Saclay, 91192 Gif-sur-Yvette, France (E-mail: merouane.debbah@ku.ac.ae). \\
\indent The work was supported by the China National Key R\&D Program under Grant 2021YFA1000500 and 2023YFB2904800, National Natural Science Foundation of China under Grant 62331023, 62101492, 62394292 and U20A20158, Zhejiang Provincial Natural Science Foundation of China under Grant LR22F010002, Zhejiang Provincial Science and Technology Plan Project under Grant 2024C01033, and Zhejiang University Global Partnership Fund, and Singapore Ministry of Education (MOE) Academic Research Fund Tier 2 MOE-T2EP50220-0019.}
}
\maketitle

\pagestyle{empty}  % no page number for the second and the later pages
\thispagestyle{empty} % no page number for the first page

\begin{abstract}
The beamforming technology with large holographic antenna arrays is one of the key enablers for the next generation of wireless systems, which can significantly improve the spectral efficiency. However, the deployment of large antenna arrays implies high algorithm complexity and resource overhead at both receiver and transmitter ends. To address this issue, advanced technologies such as artificial intelligence have been developed to reduce beamforming overhead. Intuitively, if we can implement the near-optimal beamforming only using a tiny subset of the all channel information, the overhead for channel estimation and beamforming would be reduced significantly compared with the traditional beamforming methods that usually need full channel information and the inversion of large dimensional matrix. In light of this idea, we propose a novel scheme that utilizes Wasserstein generative adversarial network with gradient penalty to infer the full beamforming matrices based on very little of channel information. Simulation results confirm that it can accomplish comparable performance with the weighted minimum mean-square error algorithm, while reducing the overhead by over 50\%.
\end{abstract}


\section{Introduction}\label{sec:intro}
The beamforming technology for massive multiple-input and multiple-output (MIMO) has been receiving significant attention due to its ability to greatly improve the throughput of communication systems. In addition, holographic MIMO communications \cite{holographic1, holographic2} have been a prospective substitute for conventional massive MIMO for its super-directivity, which is achieved by much smaller antenna element spacing and near continuous aperture. Therefore, there have been numerous researches on the design and implementation of holographic MIMO beamforming technology. However, obtaining the optimal beamforming design is usually an NP-hard problem\cite{beamformingsolution}, especially for the near-field scenario.
\par
Although classical algorithms like weighted minimum mean-square error (WMMSE)\cite{WMMSE} have good beamforming performance, they usually have huge computing and pilots overhead for holographic MIMO communication systems, due to the extensive pilot trainings and inversions of large-dimensional matrices. To address this problem, \cite{compressed} leveraged compressed sensing to improve the beamforming gain with less overhead. Besides, a Bayesian optimization-based beam alignment scheme was proposed for MIMO communication systems in \cite{Bayesian}, which exploited the information of measured beam pairs to predict the best beam direction, and achieved higher spectral efficiency only at the cost of a little overhead.
\par
\begin{figure}[t]\vspace{0mm}
	\begin{center}
		\centerline{\includegraphics[width=0.5\textwidth]{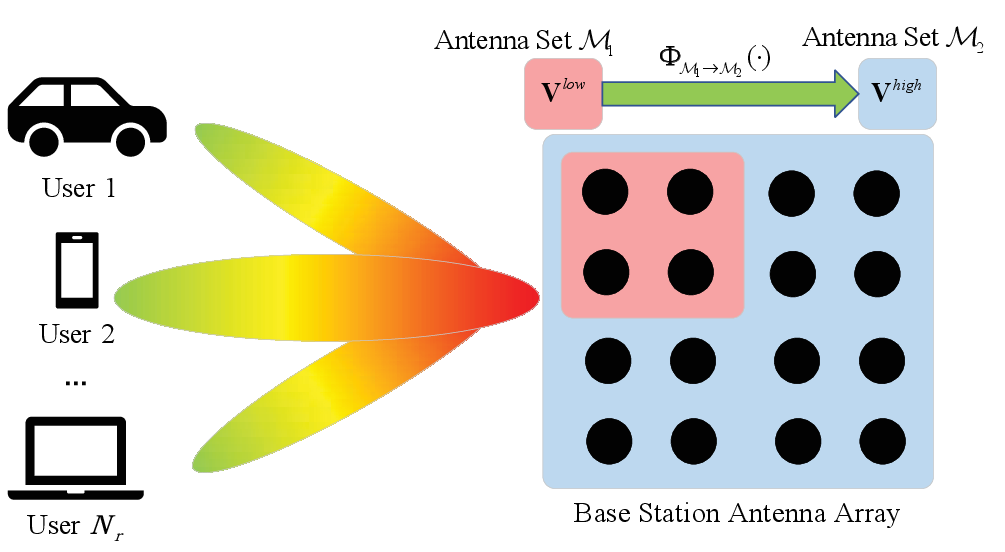}}  
		\vspace{-0mm}
        \captionsetup{font=small, name={Fig.}, labelsep=period}  
		\caption{\, The diagram shows the common system model. The function $\Phi_{\mathcal{M}_1 \rightarrow \mathcal{M}_2}(\cdot)$ transforms the low-dimensional beamforming matrix from the antenna set $\mathcal{M}_1$ to the high-dimensional beamforming matrix at $\mathcal{M}_2$.}
		\label{fig:beamforming_diagram} \vspace{-10mm}
	\end{center}
\end{figure}

Meanwhile, deep neural networks were leveraged to perform channel inferring based on the correlation between antenna elements\cite{mapping,Chafaa}, which can reduce the overhead of channel estimation as well.
% but the supervised way of learning requires a great number of labels that usually are not accessible in practical scenarios. A recent work \cite{Chafaa} successfully implemented the beam inferring across different frequencies through self-supervised deep learning. 
However, no research has been carried out on the inferring of full beamforming matrices with very small amounts of channel information. At this time as well, generative adversarial networks (GANs) were shown as powerful tools for channel estimation, signal recovery in massive MIMO communication systems, etc., since they are adept at implicitly modeling intricate probability distributions\cite{ganv2v,cgan, cwgan}. Besides, as a variant of GAN, Wasserstein generative adversarial network with gradient penalty (WGAN-GP) has been proposed in \cite{wgangp} to alleviate the problem of gradient explosion or disappearance. Jointly considering these factors motivates us to develop a GAN-based approach to obtain the high-dimensional matrix using very little channel information, as illustrated in Fig. \ref{fig:beamforming_diagram}.
 
% On the other hand, the use of artificial intelligence techniques is becoming increasingly popular in wireless communications\cite{ganv2v,wang2023energyefficient}. Among these methods, generative adversarial networks (GANs) stand out as powerful tools for unsupervised learning. They are adept at implicitly modeling intricate probability distributions and have been applied in traditional tasks like channel estimation for multi-user massive MIMO systems\cite{cgan,gan_es,cwgan}. Besides, as a variant of GAN, Wasserstein generative adversarial network with gradient penalty (WGAN-GP) has been proposed in \cite{wgangp} to alleviate the problem of gradient explosion or disappearance. Jointly considering these factors motivates us to develop a GAN-based approach to obtain the high-dimensional matrices using very little channel information, as illustrated in Fig. \ref{fig:beamforming_diagram}.

\vspace{-0mm}

\begin{figure*}[t]\vspace{-1mm}
	\begin{center}
		\centerline{\includegraphics[width=0.9\textwidth]{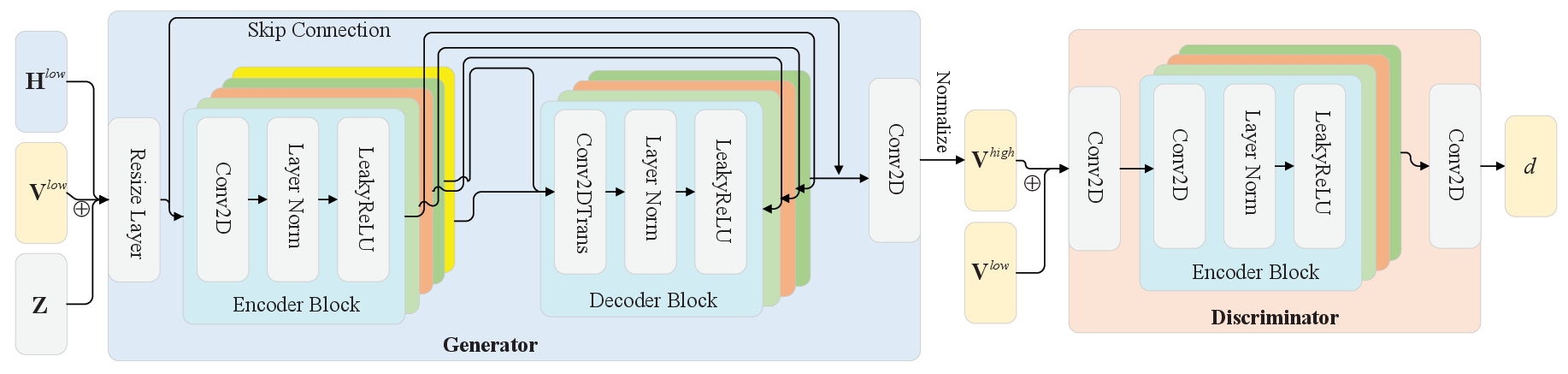} } \vspace{-2mm}
		\captionsetup{font=small, name={Fig.}, labelsep=period}  
		\caption{\, Network architecture of the proposed conditional WGAN-GP approach.}
		\label{fig:framework}
      \vspace{-10mm}
	\end{center}
\end{figure*}
\par
In this paper, we present a beamforming inferring scheme for holographic antenna arrays based on conditional WGAN-GP. Unlike previous works that have focused mainly on building correlations between channels, the proposed scheme infers high-dimensional beamforming matrices from extremely low-dimensional channel information, which exploits the ability of distribution extraction in WGAN-GP without requiring the complex design of neural network and loss function. The generative model (i.e. generator) and the adversarial model (i.e. discriminator) are trained adversarially to provide an adaptive loss. Specifically, the parameter of the generator is not updated directly from the real data, but from the backpropagation of the discriminator. This helps the generator learn the exact distribution of the target beamforming matrices. Moreover, simulation results show that the proposed conditional WGAN-GP scheme can efficiently derive the high-dimensional beamforming matrices by decreasing the overhead over 50\% compared with the traditional schemes.

\section{System Model}\label{sec:format}
In this section, we first introduce the system model for the considered communication scenario, as shown in Fig.  \ref{fig:beamforming_diagram}, and then provide the target optimization problem. Specifically, we consider a downlink millimeter wave (mmWave) communication system\cite{zhu2023robust}, where there is a base station (BS) equipped with an array of $N_t$ antenna elements, and $N_{RF}$ radio frequency chains. Besides, we assume full-digital beamforming at BS. On the receiver side, there are $N_r$ single-antenna mobile users. We denote the transmitted symbol for user $k$ as $x_k \in \mathbb{C}$, where the average energy is $\mathbb{E}\left\{ |{x}_k|\right\}= 1$. We assume a narrow-band block-fading channel model in which the $k$-th user would observe the signal as
\begin{equation}\label{Transmission Model}
y_k =\mathbf{h}_k^H \mathbf{v}_k {x}_k + \sum_{j \neq k }\mathbf {h}_k^{H}\mathbf{v}_j {x}_j + n,
\end{equation}
where $\mathbf{h}_k$ is an $ N_t \times 1$ vector that represents the mmWave channel between the BS and the $k$-th user, $\mathbf{H} = [\mathbf{h}_1, \mathbf{h}_2, \ldots, \mathbf{h}_{N_r}]$ is the channel of all users, and $\mathbf{V} = [\mathbf{v}_1, \mathbf{v}_2, \dots,\mathbf{v}_{N_r}] \in \mathbb C^{N_t \times N_r}$ is the beamforming matrix. The additive noise $n$ satisfies the circularly symmetric complex Gaussian distribution with zero mean and variance $\sigma^{2}$.
\par
As for the channel model, we would adopt the WAIR-D dataset\cite{WAIR-D}, in which the time-domain channel matrix $\mathbf h_{d}$ consists of $L$ channel paths, and each path has a time delay $\tau_{l} \in \mathbb R$, and a pair of azimuth/elevation angles of arrival (AoA) $\theta_{l}, \varphi_{l}$. Let $\rho$ denote the path loss between the user and the BS, and $p(\tau)$ represents a pulse shaping function for $T_{S}$-spaced signaling evaluated at $\tau$ seconds. Thus, the channel vector can be formulated as
\begin{equation}\label{DeepMIMO Channel Model}
\mathbf h_{d}= \sqrt \frac{M}{\rho}\sum_{l=1}^{L}\alpha_{l} p(dT_{S}-\tau_{l})  \mathbf a(\theta_{l}, \varphi_{l}),
\end{equation}
where $\mathbf a(\theta_{l}, \varphi_{l})$ is the array response vector of the BS at the AoA
$\theta_{l}, \varphi_{l}$. Therefore, the $k$-th user can obtain the spectral efficiency as follows,

\begin{equation}\label{Spectral Efficiency}
R_k(\mathbf{V})
= \log_2 (1+\frac{|\mathbf {h}_k^{H}\mathbf{v}_k|^{2}}{\sigma^{2}+\sum_{j \neq k }|\mathbf {h}_k^{H}\mathbf{v}_j|^{2}}).
\end{equation}
Here all the beamforming matrices satisfy the power constraint $\mathrm{Tr}(\mathbf{V}^H\mathbf{V}) \leq P$, and $\mathrm{Tr(\cdot)}$ denotes the calculation of the matrix trace. The signal-noise ratio (SNR) can be defined as $10\log(P/\sigma^{2})$.
\par
The main idea of this paper is to focus on how the generator $\mathcal G$ and discriminator $\mathcal D$ can be jointly optimized to infer the high-dimensional beamforming matrix $\mathbf {V}$ (i.e. $\mathbf {V}^{high}$) from low-dimensional channel $\mathbf{H}^{low} = [\mathbf{h}_1^{low}, \mathbf{h}_2^{low}, \ldots, \mathbf{h}_{N_r}^{low}]$ and low-dimensional beamforming matrix $\mathbf{V}^{low} = [\mathbf{v}_1^{low}, \mathbf{v}_2^{low}, \ldots, \mathbf{v}_{N_r}^{low}]$. Therefore, the optimization problem can be written as
\begin{equation}\label{Target Spectral Efficiency}
\begin{split}
    &\mathop {\rm max}\limits_{ \mathcal G, \mathcal D}  \mathbb{E}_{\mathbf h,P}[\sum_k  R_k({\mathbf{V}^{high}})]\\
    	&\mathrm{s.t.\ \ } \mathrm{Tr}((\mathbf {V}^{high})^H  \mathbf {V}^{high}) \leq P,
\end{split}
\end{equation}
where $\mathbb{E}_{\mathbf h,P}[ \cdot ]$ represents the average over channel samples and the sum power constraints. 

\section{Beamforming Inferring via Conditional WGAN-GP}\label{sec:beamforming_mapping}
In this section, we present the generative adversarial deep learning architecture, and discuss how to train and apply this architecture in practical scenarios. The detailed network structure is presented in Fig$.$~\ref{fig:framework}, and it is mainly composed of two parts: the generator and the discriminator, which are presented in the following subsections.

\subsection{The Network Architecture}\label{architecture}
We treat $\mathbf{V}^{low}$ as two-channel images with dimensions $ N_t^{low} \times N_r \times 2$, where the factor 2 is the result by separately considering of the real and imaginary parts. Then, we denote the random noise as $\mathbf{Z} \in \mathbb{R}^{ N_t^{low} \times N_r \times 2}$, and subsequently input these data into the generator.
\par
The generator in our model consists of four main components: a resize layer, an encoder group, a decoder group, and an independent conv2d layer. The resize layer includes two decoder blocks and one encoder block to reshape the input variables. Following the resize layer, the architecture branches into the encoder group and the decoder group, respectively. The encoder group consists of five encoder blocks, whereas the decoder group is composed of four decoder blocks. The primary functions of the encoder and decoder blocks are to downsample and upsample the data, respectively. In detail, the encoder block utilizes a sequence of conv2d, layer normalization, and leakyrelu layers to effectively extract features. Conversely, the decoder block employs transpose convolutions in place of conv2d to facilitate image reconstruction. Furthermore, the design integrates skip connections between the encoder and decoder groups. These connections are pivotal in enhancing the robustness of the model, allowing for more efficient and effective information flow during the generative process. After the final conv2d layer in the generator, the generated beamforming matrix $\mathbf{\widetilde{V}}^{high}_{k}$ must be normalized to satisfy the power constraint in (\ref{Target Spectral Efficiency}), which is expressed as follows
\begin{equation} \label{vector normalization}
    \mathbf{v}^{high}_{k} = \sqrt{\frac{P}{\sum\limits_{ 1 \leq j \leq N_r} | \mathbf{(\widetilde{v}}^{high}_{j})^H\mathbf{\widetilde{v}}^{high}_{j}}|^2} \mathbf{\widetilde{v}}^{high}_{k}.
\end{equation}
The high-dimensional beamforming matrix $\mathbf{V}^{high} = [\mathbf{v}_1^{high}, \mathbf{v}_2^{high}, \ldots, \mathbf{v}_{N_r}^{high}]$ is the ultimate output of the generator. During training, it gradually approaches the ground truth of the high-dimensional beamforming matrix $\mathbf{V}^{real}$, which is generated from the high-dimensional channel data $\mathbf{H}^{real}$ using WMMSE.
\par
The generated high-dimensional beamforming matrix $\mathbf{V}^{high}$ is concatenated with $\mathbf{V}^{low}$ and sent to the discriminator, which is composed of a conv2d layer, four encoder blocks and a conv2d layer in sequence. The discriminator is designed to assess how the generator data is close to the distribution of real data, whose output $d$ is Wasserstein distance rather than a true probability. Considering that the Kullback-Leibler divergence or Jensen-Shannon divergence in traditional GAN would change abruptly, it might result in unstable gradient descent in training process. Therefore, we utilize the Wasserstein distance instead. This distance is smooth\cite{pmlr-wgan} and can provide necessary gradient descent even when the generated distribution has no overlap with the true distribution. Thus, the conditional WGAN-GP loss with the generator $\mathcal{G}$ and the discriminator $\mathcal{D}$ can be defined as
\begin{equation}\label{generator loss}
\begin{split}
 \mathcal{L}_1 & =  \mathbb{E}_{\mathbf{V}^{high} \sim p_{real}}[\mathcal{D}(\mathbf{V}^{high}|\mathbf{V}^{low}; \theta_d )] \\
& - \mathbb{E}_{\mathbf{Z} \sim p_{z}}[\mathcal{D}(\mathcal{G}(\mathbf{Z}|\mathbf{V}^{low}, \mathbf{H}^{low}; \theta_g );\theta_d)] - \lambda GP(\mathcal{D}).
\end{split}
\end{equation}
where $\lambda$ is the penalty coefficient; $\mathbb E$ determines the value of expectation, and the gradient penalty item
\begin{equation}\label{gradient penalty}
\begin{split}
GP(\mathcal{D}) & = \mathbb{E}_{\mathbf{\widetilde V}^{high} \sim p_{sample}}[(\| \nabla_{\mathbf{\widetilde V}^{high}}\mathcal{D}(\mathbf{\widetilde V}^{high}| \\
& \mathbf{V}^{low}; \theta_d )\|_2 - 1)^2].
\end{split}
\end{equation}
In the above (\ref{generator loss}) and (\ref{gradient penalty}), $\| \cdot \|$ is the $L_2$ norm of the matrix. $\theta_g$ and $\theta_d$ denote parameters of generator and discriminator respectively, and the distribution $p_{sample}$ samples uniformly along straight lines between pairs of points conforming to the real data distribution $p_{real}$ and the generated distribution $p_{g}$. The sample $\mathbf{\widetilde V}^{high}$ is obtained by adding two parts, one of which is obtained by multiplying a random number $\alpha \sim U(0,1)$ by a real sample $\mathbf{V}^{high} \sim p_{real}$, and the other is obtained by multiplying $(1-\alpha)$ by a mimic sample $\mathbf{V}^{high} \sim p_{g}$. The input to the generator and discriminator networks are both conditioned on the information $\mathbf{V}^{low}$, which controls how the target $\mathbf{V}^{high}$ is generated. In addition, to ensure the right direction of convergence, an $\mathcal{L}_2$ loss is added to (\ref{generator loss}), expressed as
\begin{equation}
     \mathcal{L}_2  =  \mathbb{E}_{\mathbf{V}^{high} \sim p_{real}}[\| \mathcal{G}(\mathbf{Z}|\mathbf{V}^{low}, \mathbf{H}^{low}; \theta_g ) - \mathbf{V}^{high}\|_2^2]. 
\end{equation}
Finally, the objective function is the combination $\mathcal{L}_1$ and $\beta\mathcal{L}_2$
\begin{equation}
     \mathop{\rm min}\limits_{\theta_g} \mathop{\rm max}\limits_{\theta_d} \mathcal{L}_1 +\beta \mathcal{L}_2,
\end{equation}
where $\beta$ is a hyper-parameter used to scale the loss $\mathcal{L}_2$.
\subsection{Training and Prediction}
\begin{figure}[t]
	\begin{center}
		\centerline{\includegraphics[width=0.5\textwidth]{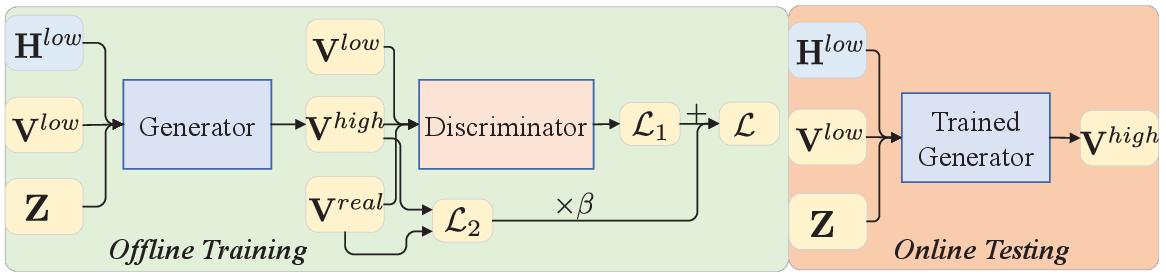}}  \vspace{-0mm}
		\captionsetup{font=small, name={Fig.}, labelsep=period}
		\caption{\, Training and prediction stages.}
		\label{fig:cwgan-gp-deploy} \vspace{-10mm}
	\end{center}
\end{figure}

In this section, we leverage the network architecture described in Section \ref{architecture} to reduce the overhead in channel estimation and beamforming, as shown in Fig$.$~\ref{fig:cwgan-gp-deploy}. The implementation of the proposed conditional WGAN-GP involves two stages: training and prediction. 
\par
\textbf{Training stage:} First, complete channel estimation is conducted to acquire the high-dimensional $\mathbf{H}^{real}$ and the low-dimensional $\mathbf{H}^{low}$. Next, WMMSE beamforming is executed to obtain the respective beamforming matrices, $\mathbf{V}^{real}$ and $\mathbf{V}^{low}$. After that, the low-dimensional matrices, $\mathbf{H}^{low}$ and $\mathbf{V}^{low}$, are combined with stochastic noise $\mathbf{Z}$, generating the high-dimensional $\mathbf{V}^{high}$ by the generator. Then, the loss function $\mathcal{L}_1$ and $\mathcal{L}_2$ are calculated and added, based on which backpropagation is employed to optimize the generator and discriminator. Once enough epochs have been executed, the proposed conditional WGAN-GP has the ability to infer the high-dimensional beamforming matrices from the low-dimensional beamforming matrices.
\par
\textbf{Prediction stage:} This stage only requires very little channel information from a small number of antennas, obviating the requirement for full channel estimation. Then, the low-dimensional beamforming matrix $\mathbf{V}^{low}$ is obtained using WMMSE. Afterwards, the trained generator directly generates the high-dimensional beamforming matrix $\mathbf{V}^{high}$ and applies it to optimize spectral efficiency in (\ref{Spectral Efficiency}). This stage only requires very little channel information from a small number of antennas, avoiding a large number of pilot trainings and inversions of high-dimensional matrix operations during channel estimation and beamforming. \textcolor{black}{Furthermore, in practical dynamic environments, transfer learning can be employed to accelerate the adaptation to the channel.}

\vspace{-1mm}

\begin{figure}[t]
	\begin{center}
		\centerline{\includegraphics[width=0.75\linewidth]{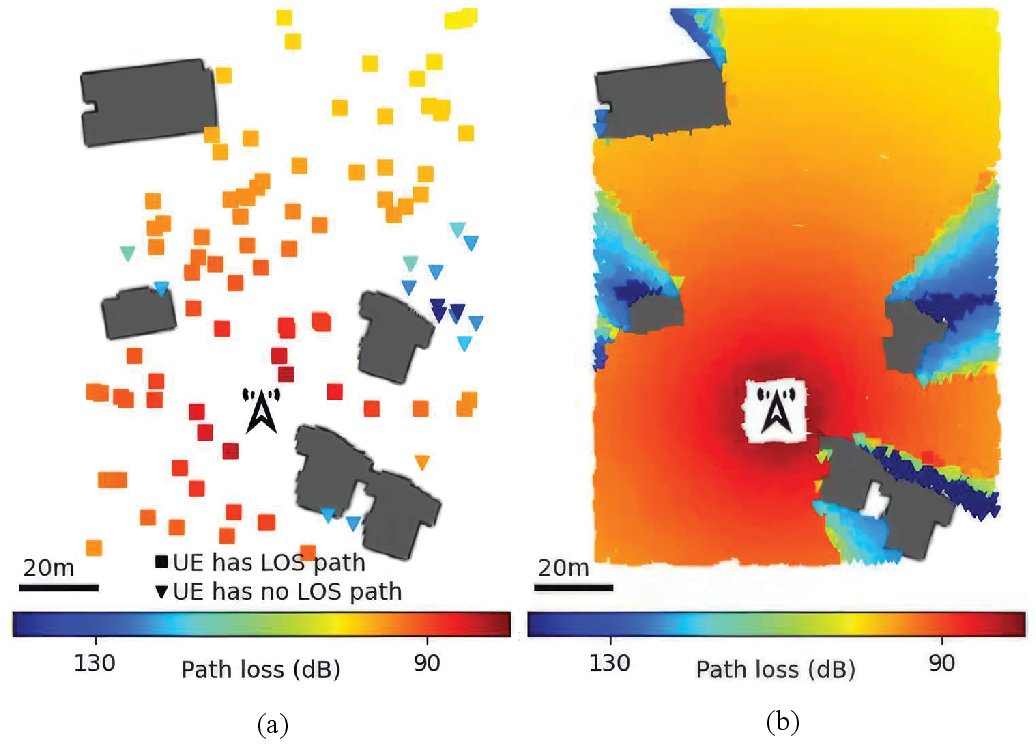}}  \vspace{-0mm}
		\captionsetup{font=small, name={Fig.}, labelsep=period}  
		\caption{\, Top view of the scenario-2 case 1 of the WAIR-D. (a) shows 100 users among the total 10000 users, (b) shows all the 10000 users.}
		\label{fig:wair-d_map} \vspace{-10mm}
	\end{center}
\end{figure}

%%% Simulation part
\section{Performance Evaluation Results}\label{sec:simulation}
In this section, we first discuss the selection of the channel dataset, and then describe the simulation parameters. Finally, we present the simulation results assessing the performance of the proposed beamforming inferring scheme.
% \vspace{-4mm}
\subsection{Simulation Settings}
To obtain high-quality deep learning models, it is necessary to have high-quality data sources. DeepMIMO\cite{alkhateeb2019deepmimo} is a well-known and widely used open-source dataset in wireless communications, constructed using 3D ray tracing and virtual scene generation technologies. Although DeepMIMO provides a large number of scenarios, its limitation lies in its reliance on virtual-generated scenes, which may not accurately simulate real communication scenarios. Additionally, the dataset only provides one setting for each scene type, making the trained model incapable of adapting to varied environments. In contrast, the recently released open-source dataset WAIR-D \cite{WAIR-D} makes a significant improvement addressing this issue. \textcolor{black}{While DeepMIMO is synthesized by simulation software modeling, WAIR-D is based on multiple practical scenario maps,} differing not only in data generation but also in storage methods. \textcolor{black}{Employing WAIR-D can lead to simulations that are more accurate in actual scenarios}. Details of the simulation parameters for the WAIR-D dataset can be found in Table \ref{systemparameter}.
\par
To better tailor the communication model to our needs, we choose scenario-2 in WAIR-D for our experiment, as it includes up to 10000 users and one BS in each case. And $N_r$ is set to 4 in the simulations. The details of the training map are presented in Fig.~\ref{fig:wair-d_map}. Antenna number 1 and 2 refer to the number of antennas in antenna sets $\mathcal{M}_1$ and $\mathcal{M}_2$, respectively. As part of the training and evaluation process, we randomly select 250 samples from the dataset and split them into two subsets with a 4:1 ratio. \textcolor{black}{The input channel matrix $\mathbf{H}$ is normalized, and the normalized channel $\widetilde{\mathbf{H}}$ can be expressed as $\widetilde{\mathbf{H}} = \mathbf{H}/\mathrm{max}(\mathrm{abs}(\mathbf{H}))$, where $\mathrm{max}$ and $\mathrm{abs}$ are maximum and absolute value operators, respectively.
In order to match the practical scenario, the channel estimation error (CEE) is set to -20dB, which is expressed as $\mathrm{CEE} = 10\log_{10}\left[\frac{\mathbb{E}[\| \mathbf{H} -  \hat{\mathbf{H}} \|_2^2]}{\mathbb{E}[\| \mathbf{H} \|_2^2]}\right]$, where $\hat{\mathbf{H}}$ is the estimated channel matrix.}
\par
The proposed model is trained by applying the RMSProp algorithm with a learning rate of $2 \times 10^{-4}$ and $2 \times 10^{-5}$ to the generator and discriminator, respectively. Additionally, $\beta$ is set to 100, and the parameters of the generator and the discriminator are updated every 1 and 5 batches, respectively. The model is trained for 50 epochs with a batch size of 1 using TensorFlow 2.8 on Ubuntu 20.04. The computer used for training includes an EPYC 75F3 CPU and an RTX 3090 GPU. The normalized mean-squared-error (NMSE) is used to quantify the difference between the estimated beamforming matrix $\mathbf{V}^{high}$, and the target beamforming matrix $\mathbf{V}^{real}$.
It is expressed as $\rm{NMSE} = 10 \log_{10}{\mathbb{E}\left[\frac{\| \mathbf{V}^{real} - \mathbf{V}^{high} \|^2
_2}{\| \mathbf{V}^{real} \|_2^2}\right]}$.
% NMSE values are obtained in decibels by calculating $10\log_{10}\left(\cdot \right)$.
\begin{table}[t]
\centering
\caption{System Parameters}\vspace{-2mm}
\label{systemparameter}
\begin{tabular}{cccc}
 \toprule
\textbf{Parameters} & \textbf{WAIR-D} & \textbf{Parameters} & \textbf{WAIR-D} \\
 \midrule
 Base Station         & BS0    &        Radio Chain Number    & 32  \\
 Antenna Number 2  & 32     & Antenna Spacing         & $\lambda/2$ \\
 Antenna Number 1 & 8      & Path Number          & 5 \\
 Carrier Frequency    & 60GHz  & SNR & 10dB \\
 Bandwidth            & 50MHz  & \textcolor{black}{CEE} & \textcolor{black}{-20dB} \\
 \bottomrule
\end{tabular}
\end{table}\vspace{-2mm}

% \begin{table}[h!]\vspace{-4mm}
% \centering
% \caption{System Parameters}
% \label{systemparameter}
% \begin{tabular}{cc}
%  \toprule
%  Parameters           & WAIR-D  \\
%  \midrule
%  Base Station          & BS0 \\
%  Antenna Number 0     & 32  \\
%  Antenna Number 1     & 8   \\
%  Carriar Frequency    & 60GHz  \\ 
%  Antenna Spacing      & $\lambda/2$  \\
%  Radio Chains         & 1     \\ 
%  Signal Power         & 1dBm  \\ 
%  Path Number          & 5   \\
%  Bandwidth            & 50MHz \\
%  PNR                  & 20dB \\ 
%  \bottomrule
% \end{tabular}
% \end{table}

\begin{figure}[ht]\vspace{0mm}
	\begin{center}
		\centerline{\includegraphics[width=0.40\textwidth]{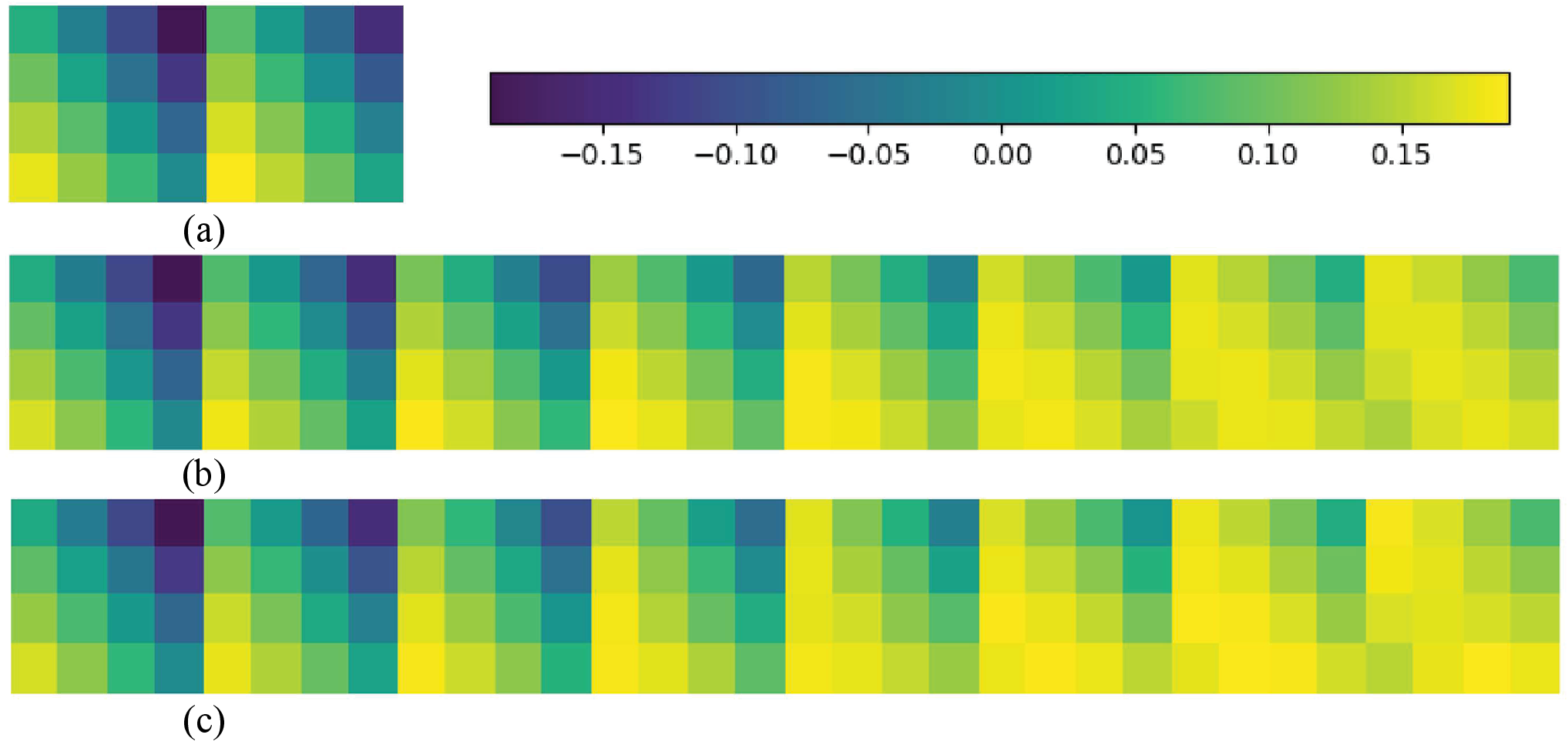} } \vspace{-0mm}
		\captionsetup{font=small, name={Fig.}, labelsep=period}  
		\caption{\, Pseudo-color plots for real components of (a) low-dimensional beamforming matrix $\textcolor{black}{\mathbf{V}^{low}}$, (b) target high-dimensional beamforming matrix $\textcolor{black}{\mathbf{V}^{real}}$ and (c) predicted high-dimensional beamforming matrix $\textcolor{black}{\mathbf{V}^{high}}$.}
		\label{fig:mapping_sample}
      \vspace{-10mm}
	\end{center}
\end{figure}

\subsection{Simulation Results}
In Fig$.$~\ref{fig:mapping_sample}, we demonstrate the proposed inferring scheme's performance by visualizing the beamforming matrices. Subplots (a), (b) and (c) display the real part of the input, target, and predicted matrix, respectively. 
The horizontal axis represents the antenna index, while the vertical axis corresponds to the user index. Each colored block represents a real number, with larger absolute values indicating a stronger response from the respective antenna.
Comparing  Fig$.$~\ref{fig:mapping_sample}(a) and Fig$.$~\ref{fig:mapping_sample}(c), it can be seen that the low-dimensional matrix is a part of the high-dimensional one. This suggests that the inferring process enlarges the dimension of the beamforming matrix while preserving its original components. As shown in Fig$.$~\ref{fig:mapping_sample}(b) and Fig$.$~\ref{fig:mapping_sample}(c), there is a notable similarity between the ground truth beamforming matrix and the beamforming matrix produced by the conditional WGAN-GP, highlighting the detailed and accurate nature of the inferring method. 
\par
\begin{figure}[t]\vspace{0mm}
	\begin{center}
		\centerline{\includegraphics[width=0.35\textwidth]{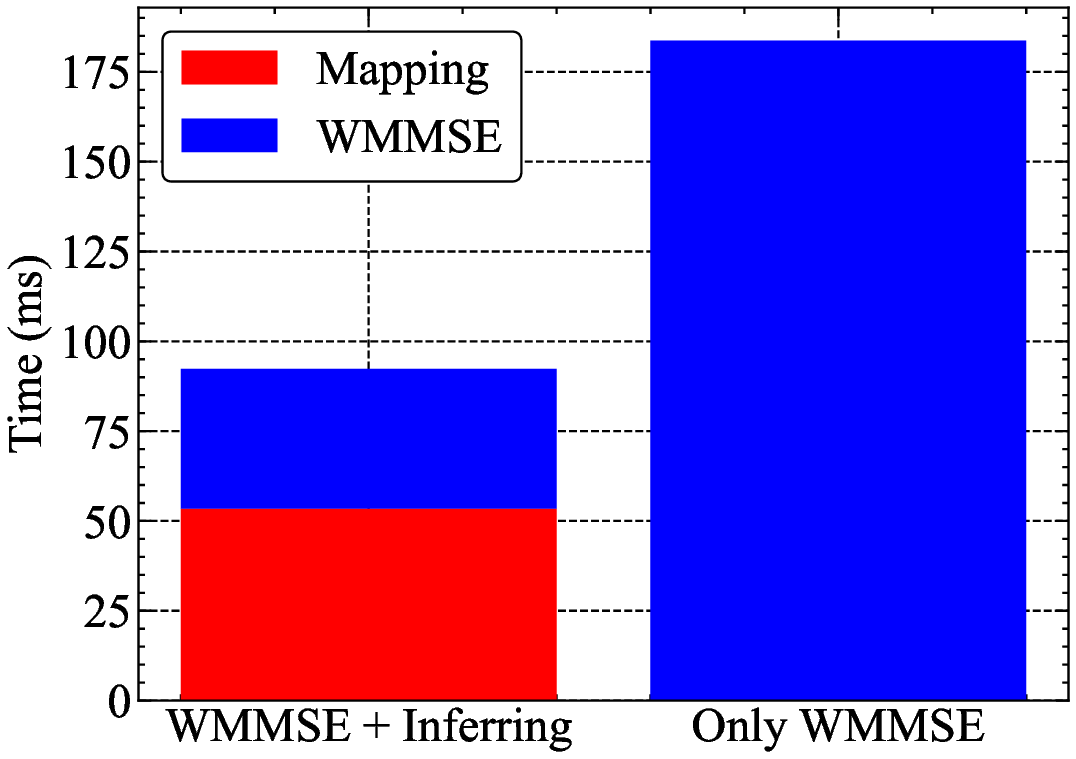} } \vspace{-0mm}
		\captionsetup{font=small, name={Fig.}, labelsep=period}  
		\caption{\, Comparison of running time between the original WMMSE algorithm and the proposed algorithm.}
		\label{fig:compare}
      \vspace{-8mm}
	\end{center}
\end{figure}

In Fig$.$~\ref{fig:compare}, a comparison is made between the running time of the proposed algorithm and the traditional WMMSE algorithm. \textcolor{black}{The proposed scheme is trained on GPU. To ensure fairness, both algorithms are executed only on the CPU while the stopping criterion $\epsilon$ in the WMMSE algorithm is set to $10^{-4}$.} The results demonstrate that inferring high-dimensional beamforming matrices from low-dimensional ones reduces the time required by over 50\%. This can be attributed to the generator’s ability in generating and inferring between different distributions whilst taking into account the WMMSE algorithm that requires multiple iterations without utilizing historical data.
% \textcolor{black}{Furthermore, in practical deployment scenarios, the proposed scheme saves even more time as the neural network's processing speed can be increased through common accelerators such as GPUs, while the traditional WMMSE algorithm can only be run on CPUs, which shows the high potential and efficiency of the proposed scheme.}
Furthermore, the complexity of the original WMMSE algorithm and the proposed algorithm could be expressed as $O(L(N_t^{high})^3)$ and $O(L(N_t^{low})^3) + O(N_r(N_t^{high})^2+(N_t^{high})^3)$, where $L$ indicates the number of WMMSE iterations, while $N_t^{low}$ and $N_t^{high}$ denote the dimension of low-dimensional and high-dimensional beamforming matrices, respectively. The inferring scheme shows a considerable advantage over the traditional WMMSE algorithm as $N_t^{low}$ is significantly smaller than $N_t^{high}$, and $N_r$ is much smaller than $LN_t^{high}$.
% These findings confirm the previous simulation results.
These findings are confirmed by the previous simulation results.
% \begin{table}[h!] \vspace{-1mm} 
% \centering
% \caption{Comparison of Complexity}
% \label{complexity}
% \begin{tabular}{|c c|}
%  \hline
%  Methods  & Test Complexity  \\ [1.0ex]
%  \hline\hline
%  Only WMMSE   & \multicolumn{1}{c|}{$O(L(N_t^{high})^3)$} \\
% WMMSE+Mapping & \multicolumn{1}{c|}{$O(L(N_t^{low})^3) + O(N_r(N_t^{high})^2+(N_t^{high})^3))$}  \\
%  \hline
% \end{tabular}
% \end{table} \vspace{-1mm}

  \vspace{-0mm}

\begin{figure}[t]\vspace{-0mm}
	\begin{center}
		\centerline{\includegraphics[width=0.35\textwidth]{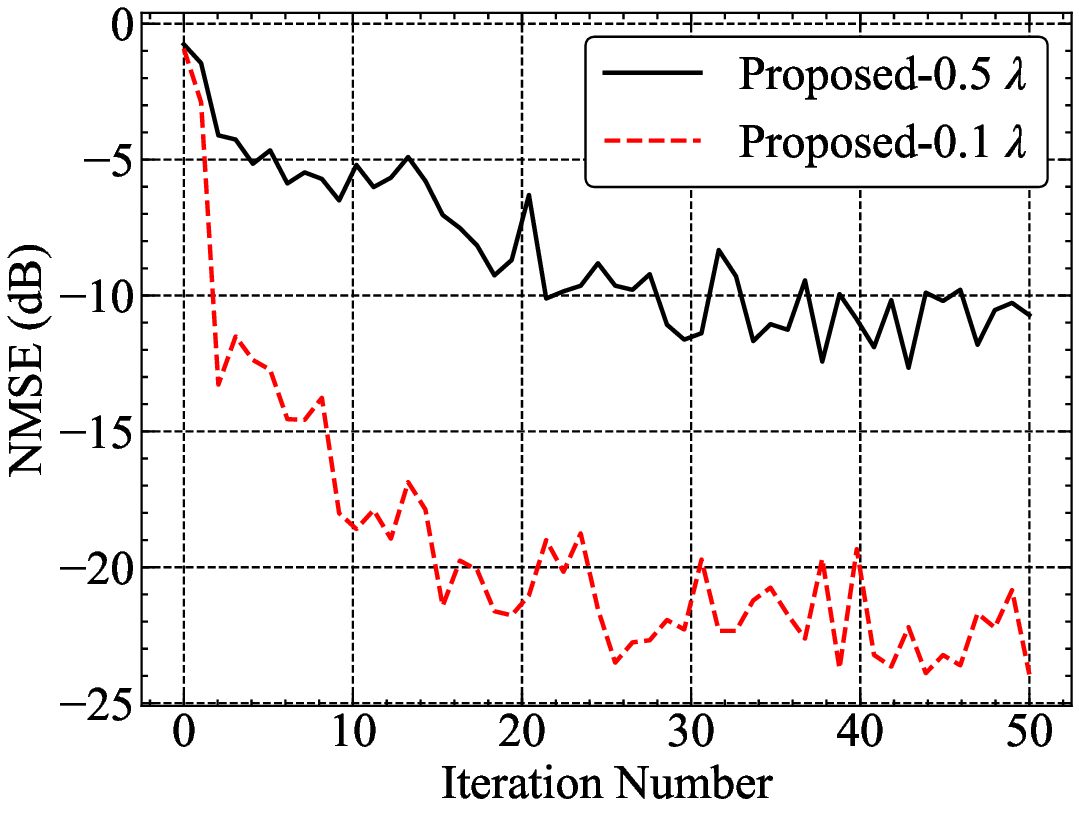} } \vspace{-0mm}
		\captionsetup{font=small, name={Fig.}, labelsep=period}  
		\caption{\, Comparison of NMSE between holographic and conventional antenna arrays with different number of iterations.}
		\label{fig:nmse}
      \vspace{-10mm}
	\end{center}
\end{figure}\vspace{0mm}

\begin{figure}[t]\vspace{0mm}
	\begin{center}
		\centerline{\includegraphics[width=0.35\textwidth]{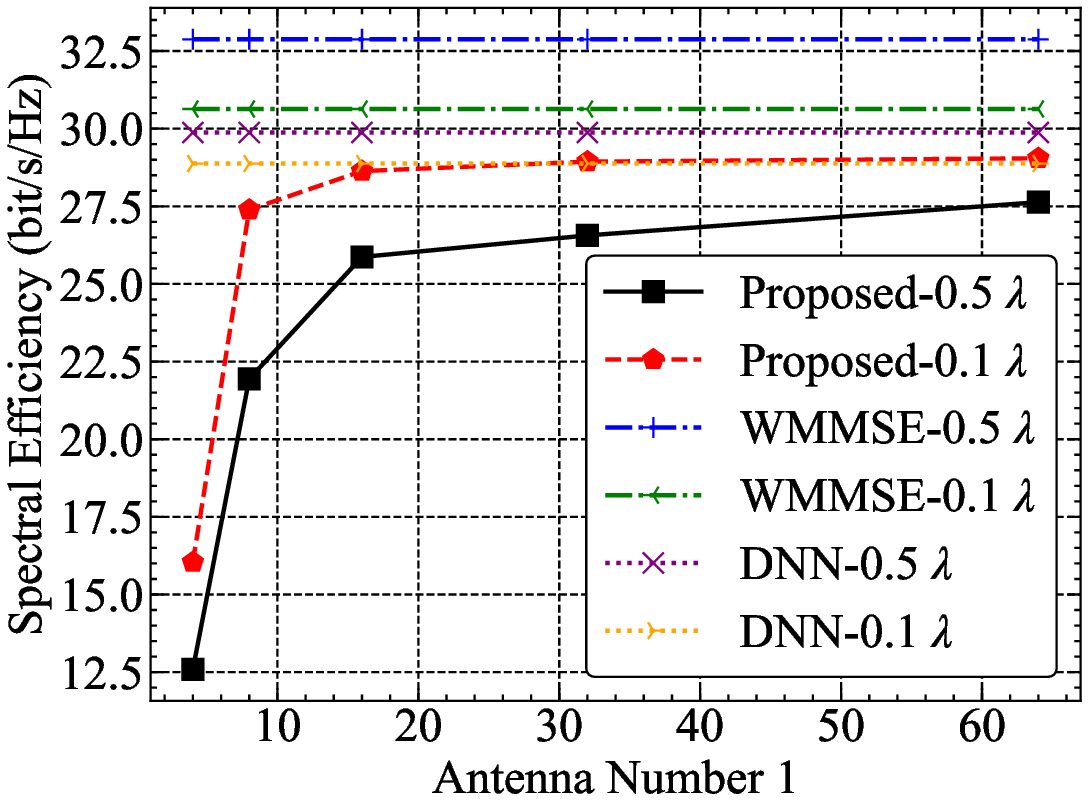} } \vspace{-0mm}
		\captionsetup{font=small, name={Fig.}, labelsep=period}  
		\caption{\, Spectral efficiency with antenna number 1.}
		\label{fig:se}
      \vspace{-10mm}
	\end{center}
\end{figure}\vspace{0mm}
\par
In Fig$.$~\ref{fig:nmse} and Fig$.$~\ref{fig:se}, we evaluate the performance of the proposed inferring technique at different antenna spacing and antenna number 1, where the $0.1\lambda$ case refers to beamforming with holographic antenna arrays. It is evident from Fig$.$~\ref{fig:nmse} that there is a gradual reduction in NMSE with an increase in the iteration numbers, and the utilization of smaller antenna spacing can lead to a faster descent speed and lower NMSE. Correspondingly, Fig$.$~\ref{fig:se} shows the system sum spectral efficiency of the proposed method and the WMMSE as well as the DNN \cite{zhu2023robust} method as antenna number 1 increases, while keeping the antenna number 2 fixed at 64. It is demonstrated that the performance of WMMSE and DNN degrades as the antenna spacing decreases from $0.5\lambda$ to $0.1\lambda$, which can be attributed to the smaller aperture and lower resolution. On the contrary, the performance of the proposed method increases with smaller antenna spacing, but still bounded by the WMMSE method. Moreover, the smaller antenna spacing can lead to a faster ascent speed and higher spectral efficiency. This phenomenon can be explained by the high spatial correlation as a result of such a smaller antenna spacing, which helps the conditional WGAN-GP understand the underlying channel distribution of adjacent antennas more effectively. It is noted that the proposed scheme can attain comparable performance with the DNN and the WMMSE method while requiring significantly less CSI, which greatly reduces the overhead of channel estimation.

% This also further reflects why our proposed inferring scheme can achieve competitive performance with the classical WMMSE algorithm.

\section{Conclusion}\label{sec:conclusion}
In this paper, a beamforming inferring scheme was proposed for holographic antenna arrays using conditional WGAN-GP, where a generator and a discriminator were designed with a composite loss function using the Wasserstein distance and the $L_2$ norm. Specifically, the generator can learn the ground truth distribution of the target beamforming matrices through adversarial training. Extensive numerical simulations show that the proposed scheme can not only achieve competitive performance compared with the conventional WMMSE algorithm, but also can reduce resource overhead significantly. In addition, it also indicates that the smaller antenna spacing can lead to a faster ascent speed and higher spectral efficiency, which further reveals the application prospects of holographic antenna arrays. Future work will focus on extending the inferring algorithm to the MIMO wideband channel scenarios and considering the incorporation of reconfigurable intelligent surfaces to broaden the application scope.

\bibliographystyle{IEEEtran}
\bibliography{mapping}
\vspace{12pt}
\end{document}